\documentclass[12pt]{iopart}
\usepackage{epsfig}
\usepackage{graphicx}
\begin{document}

\title[Hybrid quantum devices and quantum engineering]{Hybrid quantum devices and quantum engineering}

\author{M. Wallquist$^1$, K. Hammerer$^1$, P. Rabl$^2$,  M. Lukin$^2$ and P. Zoller$^1$}

\address{$^1$Institute for Theoretical Physics, University of Innsbruck, \\
and
Institute for Quantum Optics and Quantum Information, Austrian Academy of Sciences,\\
Technikerstrasse 25, 6020 Innsbruck, Austria}
\address{$^2$ITAMP, Harvard-Smithsonian Center for Astrophysics, Cambridge, Massachusetts 02138, USA \\
and Department of Physics, Harvard University, Cambridge, Massachusetts 02138, USA}
\ead{margareta.wallquist@uibk.ac.at}
\begin{abstract}
We discuss prospects of building hybrid quantum devices involving elements of atomic and molecular physics, quantum optics and solid state elements with the attempt to combine advantages of the respective systems in compatible experimental setups. In particular, we summarize our recent work on quantum hybrid devices and briefly discuss recent ideas for quantum networks. These include interfacing of molecular quantum memory with circuit QED, and using nanomechanical elements strongly coupled to qubits represented by electronic spins, as well as single atoms or atomic ensembles.

\end{abstract}

\maketitle

\section{Introduction}

Significant progress has been made during the last few years in implementing quantum information processing in various disciplines of physics \cite{ZollerEJPD2005}. In the context of atomic, molecular and optical (AMO) physics trapped ions have demonstrated the basic building blocks of a scalable quantum computer, including high-fidelity gates between qubits \cite{SchmidtKalerNature2003,LeibfriedNature2003,BenhelmNaturePh2008} and the transport of ions in a charge-coupled device between memory and processor regions \cite{BarrettNature2004}. Light-atom interfaces and networks have been studied with single neutral atoms in high-Q cavities \cite{PinkseNature2000,McKeeverNature2003}, and with free-space atomic ensembles \cite{KuzmichNature2003,VanderWalScience2003,HammererArxiv2008,SangouardArxiv2009}. Furthermore, cold atoms and molecules in optical lattices have provided quantum simulation of strongly correlated condensed matter systems \cite{ParedesNature2004,JoerdensNature2008}. Remarkable progress has also been made on the solid state side. Examples are the recent experiments with quantum dots \cite{ElzermanNature2004,KroutvarNature2004,HennessyNature2007}, NV centers and impurities \cite{WrachtrupJPhysCM2006}, and with superconducting circuits (for a review see \cite{SpecIssueQIP,ClarkeNature2008}).
A recent highlight is circuit QED demonstrating the strong coupling of Cooper pair box qubits and Josephson phase qubits to a microwave bus provided by a superconducting stripline cavity \cite{AnsmannNature2009,FinkNature2008,SillanpaaNature2007,MajerNature2007}; for a review see \cite{SchoelkopfNature2008}.

All of the above systems have their advantages (and disadvantages). In general, AMO systems are identified as close to ideal realizations of isolated quantum systems which can be manipulated with high precision on the single quantum level, while solid state setups promise a priori scalability, benefiting directly from developments in nanotechnology. In light of the progress with AMO and solid state systems it seems timely to ask if quantum hybrid systems can be built which combine the advantages of AMO and solid state elements via a quantum interface in compatible experimental setups \cite{ZollerEJPD2005}. We note that the notion of hybrid systems is typically discussed in different contexts. While sometime taking ideas and descriptions from quantum optics with atoms to solid state physics is denoted as a hybrid approach, we will focus below on AMO solid-state hybrids and quantum interfaces where elements of atomic quantum optics are combined and interfaced with quantum solid-state devices. Hybrids have also been discussed as combining various solid-state elements, e.g. interfacing optical photonic qubits via NV centers to a microwave based quantum circuit involving Josephson qubits.

Below we will describe several examples of hybrid systems involving AMO and solid-state elements. Such hybrid systems can involve the coupling of active quantum elements, such as a solid-state quantum processor coupled to a high-fidelity atomic quantum memory; the latter could further provide an interface to optical qubits, and thus to quantum communication. Coupling AMO and solid-state elements implies the creation of a quantum interface. The interface can involve a direct coupling, which typically requires storing atoms or molecules close to solid-state surfaces - quite often associated with decoherence, and often in a cryogenic setup - which can be an experimental challenge for AMO systems. Alternatively, such quantum interfaces can be provided by an optical photon (or possibly also microwave) bus which allows to couple the systems at large(r) distances. We present examples for both types of interfaces in the following.
In section \ref{Sec:CQED} we present an AMO solid-state interface via a microwave bus consisting of a superconducting stripline cavity. Section \ref{Sec:nanomec} focuses on nanomechanical elements and their coupling to NV centers, single atoms and atomic ensembles. Finally section \ref{Sec:network} briefly presents examples of cavity QED solutions for quantum networks.

\section{Circuit QED and polar molecules}
\label{Sec:CQED}

\begin{figure}[hb]
\begin{centering}
\includegraphics[width=0.6\textwidth]{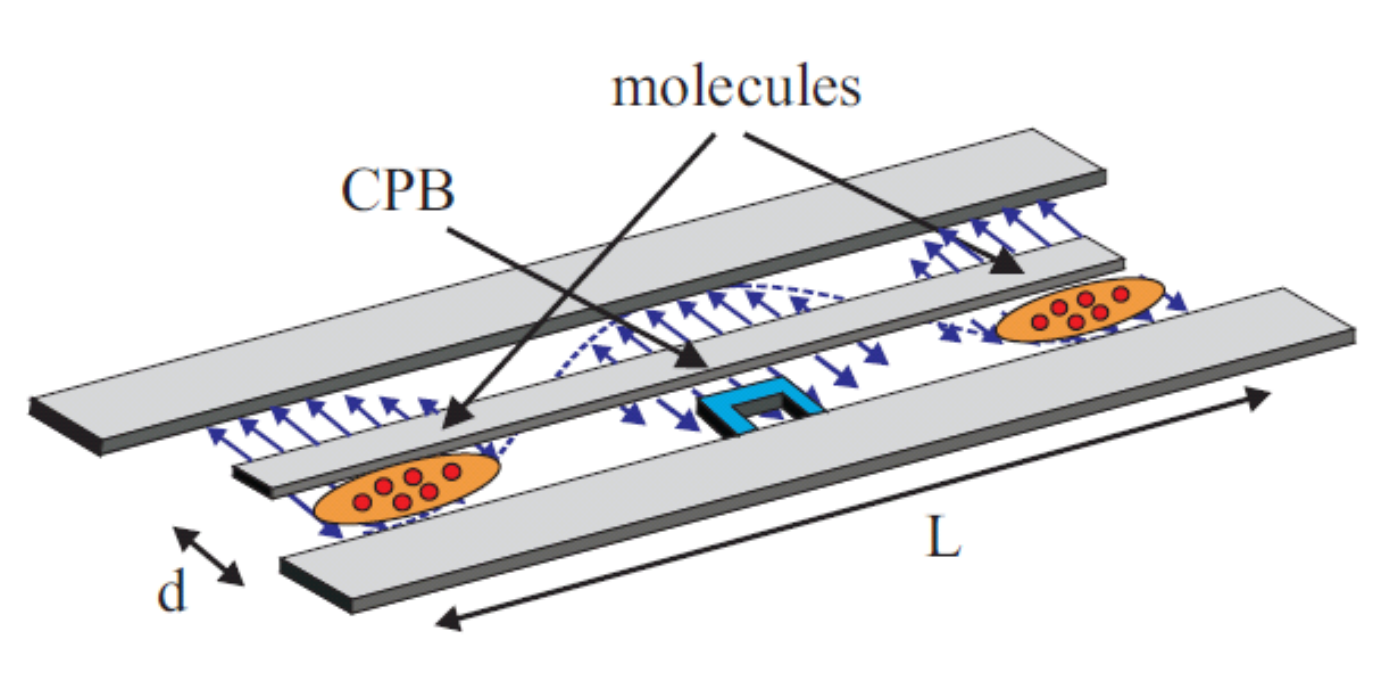}
\caption{Cooper-pair box (CPB) qubit and polar molecule ensemble couple to microwave stripline cavity photons via capacitive and dipole coupling, respectively. Strong coupling is achieved due to the quasi-1D structure of the cavity, with length L $\sim$ cm and width d $\sim \mu$m.}
\label{Fig:MolSetup}
\end{centering}
\end{figure}

As a first example of an AMO solid-state interface, we consider the coupling of polar molecules to a circuit QED setup \cite{AndreNPhys2006,RablPRL2006,RablPRA2007,YelinPRA2006}, see figure 1. The superconducting stripline cavity provides an interface for solid-state qubits to AMO systems, including Rydberg atoms \cite{SorensenPRL2004} and trapped ions \cite{TianPRL2004}, and consequently allows an optical interface for quantum communication.
We recall that the Jaynes-Cummings model can be realized with superconducting Cooper-pair box (CPB) qubits strongly coupled to microwave photons in a cm-long stripline cavity \cite{WallraffNature2004,BlaisPRA2004}; the small mode volume of its quasi-1D structure allows for a strong dipole coupling. This setup is attractive for quantum information processing, using the
stripline cavity as a high-Q photon bus for solid-state quantum processors, like in the recent implementation of the two-qubit Grover and Deutsch-Josza algorithms \cite{DiCarloNature2009}. For a typical cavity mode frequency $\omega_{\rm c} \sim 10$ GHz
and mode decay rate $\kappa \sim 10$ kHz the corresponding quality factor is $Q \sim 10^6$.
Both the cavity mode frequency and the quality factor are tunable using additional superconducting quantum interference device (SQUID) elements \cite{PalaciosJLTP2008,SandbergAPL2008}. However, the major drawback of superconducting devices for quantum information processing, and in particular for the storage of information, is their strong coupling to the environment.

With the dissipation problem in mind, the idea is to attach robust AMO memory units to the stripline cavity; here we focus on rotational states of single polar molecules or molecular ensembles.
The rotational excitations are in the microwave regime, allowing near-resonant coherent interaction with the stripline cavity photons.
Assuming a single polar molecule to be prepared in its internal vibrational and electronic ground-state, the internal dynamics is described by a rigid rotor model with anharmonic spectrum $E_N = B N (N+1)$ which interacts through electric dipole coupling with the stripline cavity photons.  The anharmonicity allows us to pick out a two-level subspace in the rotational spectrum, leaving an effective Jaynes-Cummings type Hamiltonian for the stripline-molecule system,
\[
H = \hbar \omega_{\rm c} \hat{c}^\dag \hat{c} + E_{\rm rot} \hat\sigma_z + \hbar g \left( \hat\sigma_+ \hat{c} + \hat\sigma_- \hat{c}^\dag \right)
\]
with $\hat{c}$ the cavity mode annihilation operator and $\hat\sigma_{z,\pm}$ operators for the chosen two-level rotational subspace $\{|N+1\rangle, |N\rangle \}$, and $E_{\rm rot} = (E_{N+1}-E_N)/2$. Trapping the molecular memory close to the stripline surface ($1 - 0.1 \ \mu$m), presumably using an electric Z-trap, allows a strong vacuum Rabi frequency of $g \sim 40 - 400 $ kHz \cite{AndreNPhys2006}.

With a trap frequency $\nu_t \sim 1-10$ MHz far lower than reasonable cryostat temperatures (10 mK $\sim$ 200 MHz), it is desirable to cool the motion of the trapped molecule after loading. Through the spatial dependence of the microwave driving fields, it is possible to address both the motion and the internal rotational levels, and thus to employ a technique similar to laser cooling of trapped ions \cite{StenholmRMP1986,LeibfriedRMP2003}. The low dissipation rate of rotational excitations is overcome by engineering an effective dissipation via the decay of the cavity field, allowing for  motional ground-state cooling \cite{AndreNPhys2006,WallquistNJP2008}.

The coupling of the molecular memory to the stripline cavity is strongly enhanced by replacing the single molecule with an ensemble of $N \sim 10^4 - 10^6$ polar molecules, resulting in a collectively enhanced vacuum Rabi frequency $g\sqrt{N} \sim 1 - 10$ MHz \cite{RablPRL2006}. Here the information is stored in collective excitations, which for a highly polarized ensemble are approximately described by the harmonic oscillator spectrum. The CPB qubit provides the necessary nonlinearity for performing basic quantum information operations involving the cavity and ensemble oscillators. The robustness of the ensemble memory unit is even higher in a self-assembled dipolar crystal, in which collisional dephasing is suppressed \cite{RablPRA2007}. Such a crystal is formed in the high-density limit of cold clouds of polar molecules, under 1D or 2D trapping conditions.

Collective enhancement also allows for circuit QED, despite the weaker interaction, with spin ensembles \cite{ImamogluPRL2009} or molecular ion ensembles \cite{SchusterArxiv2009} (see also \cite{PetrosyanPRL2008,TordrupPRL2008,VerduPRL2009,PetrosyanPRA2009}).

\section{Hybrid Systems involving Nanomechanical Elements}
\label{Sec:nanomec}

Lately, along with experimental progress in the fabrication and manipulation of micro- and nanomechanical systems,
a new field exploring the quantum limit of mechanical motion has emerged \cite{SchwabPhysToday2005,KippenbergOptEx2007,KippenbergScience2008,MarquardtPhys2009}. Examples include optomechanical systems, where radiation pressure of photons couples to the motion of a movable cavity-mirror \cite{GiganNature2006,ArcizetNature2006,KlecknerNature2006} or membrane \cite{ThompsonNature2008}, or nanoelectromechanical systems (NEMS) which are naturally integrated with electrical circuits such as superconducting stripline cavities \cite{RegalNaturePh2008} or single electron transistors \cite{NaikNature2006}.
However, the observation of quantum effects is challenging as it requires cooling of the mechanical motion close to the ground state  and the ability to create nonclassical states, such as squeezed, Fock or entangled states (see e.g. \cite{JaehneNJP2008,JaehnePRA2009}). Consequently, the aim is to create a coherent interface between nanomechanical degrees of freedom and a well-controlled quantum system, such as electronic spins, atoms, photons, for which there exist tools for preparation, manipulation and measurement.

\subsection{Strong magnetic coupling of electronic spin qubits to a mechanical phonon bus}

\begin{figure}[ht]
\begin{centering}
\includegraphics[width=0.55\textwidth]{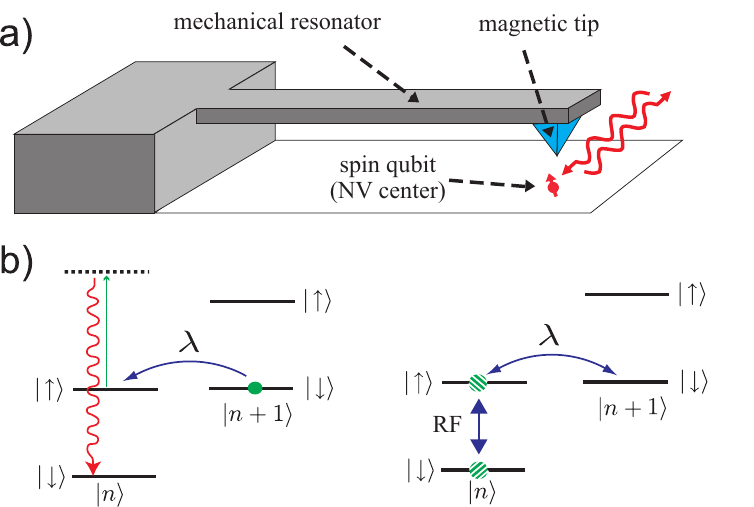}
\caption{(a) Electronic spin associated with a nitrogen-vacancy (NV) impurity in diamond couples through position dependent Zeeman shift to
motion of a NEMS with magnetized tip. (b) Cooling of the NEMS motion and creation of arbitrary resonator states using the coherent coupling to the electronic spin combined with optical pumping and RF-pulses, respectively. }
\label{fig:Transducer}
\end{centering}
\end{figure}

As a first example, we consider a setup where the quantized motion of a NEMS with magnetized tip couples directly to an isolated spin qubit through the position dependent Zeeman shift \cite{RablPRB2009}, as shown in figure \ref{fig:Transducer}(a).
For nanoscale dimensions, the resulting coupling strength $\lambda \sim 100$ kHz between the spin and a single vibrational quanta can exceed the intrinsic spin decoherence rate as well as the mechanical heating rate. The physics of this system is then well described by a Jaynes-Cummings model in the strong coupling regime, in close analogy to cavity QED.

The focus of \cite{RablPRB2009} is on a realistic implementation of such a strongly coupled system using the electronic spin associated with a nitrogen-vacancy (NV) impurity in diamond. Here spin states can be prepared and detected optically using laser excitations to electronically excited states \cite{WrachtrupJPhysCM2006}, while in the ground state spin coherence times of $1.8$ ms have been achieved even at room temperature \cite{BalasubramanianNatureMat2009}.  Ref. \cite{RablPRB2009} addresses in particular the problem of strong spin dephasing due to hyperfine interactions with the nuclear spin bath as well as the frequency mismatch of the mechanical motion (MHz) and the zero field splitting of the spin (GHz). To overcome those problems, the qubit is encoded in a dressed basis of the microwave driven NV spin which is highly insensitive to the quasi-static field of the nuclei, while at the same time a strong resonant interaction with the vibrating tip is realized.
Under such conditions the strong coupling regime between dressed spin levels and mechanical motion can be achieved under realistic conditions. Combined with the ability to optically prepare and detect the spin states, this coupling in principle enables the generation of arbitrary motional superposition states. The two basic steps are sketched in figure \ref{fig:Transducer}(b). First, a resonant transfer of a motional excitation into a spin excitation followed by optical pumping steps leads to cooling and prepares the resonator in a pure state, e.g. the motional ground state. Second, using coherent processes only, spin superpositions are converted into equivalent superpositions of motional states.  In conclusion, the strong coupling of an electronic NV impurity spin to a NEMS resonator can be used as a tool to control the mechanical motion via the spin.

The coupling of a single spin to a mechanical resonator discussed above can also be extended to a whole array of resonators~\cite{Rabl2009}. By applying a bias voltage the resonators are charged and interact with each other via long-range electrostatic forces. The magnetic interaction of the spins with the collective phonon modes of the coupled NEMS array then allows the implementation of quantum operations between distant spin qubits, in direct analogy to the well-known quantum computing proposals for trapped ions \cite{CiracPRL1995}.
This NEMS based quantum bus is applicable for a wide range of solid state spin qubits and allows the design of different spin-spin interactions and scalable quantum computing architectures by the appropriate circuit layouts \cite{Rabl2009}. Further it enables the possibility to couple dissimilar spins with each other and provides a general interface between spins and other charged based qubits, such as superconducting qubits or trapped ions.

\subsection{Strong coupling of a mechanical oscillator and a single atom}

\begin{figure}[hb]
\begin{centering}
\includegraphics[width=0.95\textwidth]{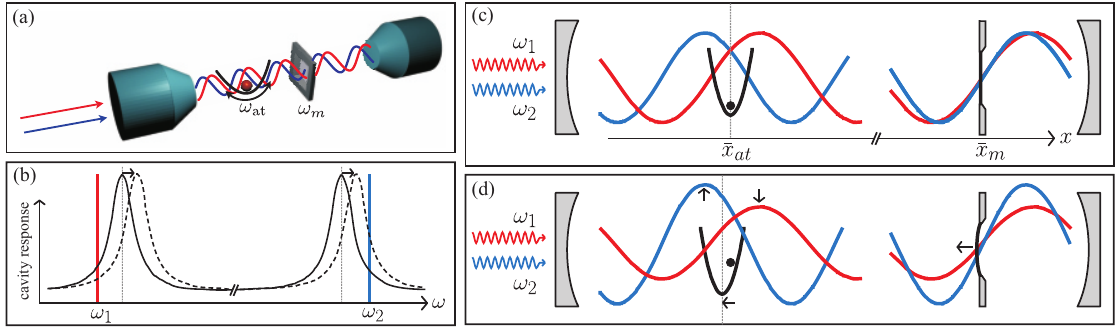}
\caption{(a) Two driven optical cavity modes mediate linear interaction between a vibrating membrane and the motion of a single trapped atom. (b) The two modes are driven on opposite sides of their respective resonances. When the membrane moves, the cavity response increases for one mode and decreases for the other. (c) When the membrane is in equilibrium, the atomic motion is centered around $\bar{x}_{at}$.
(d) When the membrane moves, the opposite response of the two fields shifts the atomic equilibrium and thus creates a linear coupling of the atomic motion to the membrane motion. }
\label{Fig:LinCoupl}
\end{centering}
\end{figure}

There are several proposals for interfacing nanomechanical elements as a novel mesoscopic system and trapped ions or atoms. One idea is to construct a mesoscopic Paul trap for ions using suspended nanomechanical resonators as the tiny trap electrodes. Thus the electrodes themselves are high-Q resonators with mechanical degrees of freedom, which can be cooled, manipulated and measured via the laser-driven ions \cite{TianPRL2004_2}. Similarly it was proposed to trap a single ion close to a voltage-biased doubly-clamped cantilever realizing a nanomechanical oscillator \cite{WinelandNIST1998,HensingerPRA2005}, or a BEC close to a cantilever with magnetic tip \cite{TreutleinPRL2007}. Also the coupling of artificial atoms to a nanomechanical resonator beam was considered, such as an embedded self-assembled quantum dot \cite{WilsonRaePRL2004} or a Cooper-pair box qubit \cite{MartinPRB2004}.

Optomechanical systems are recently approaching the quantum regime \cite{GroeblacherNature2009,GroeblacherArxiv2009,SchliesserArxiv2009}.
The dominant mechanism of these systems is the radiation pressure coupling of photons (light) to the motion of a micromirror or membrane \cite{KippenbergScience2008,MarquardtPhys2009}.
Independently, we can consider trapping an atom in the optical lattice formed by driven modes in an optical cavity.
In such a setup, the quantized motion of the atom couples to the quantized fluctuations of the light field.
The idea in \cite{HammererArxiv2009} is to combine these independent mechanisms in one and the same cavity, as shown in figure \ref{Fig:LinCoupl}(a), in order to create a cavity-mediated interaction between the atomic motion and the membrane motion.

For a strongly driven cavity mode $\hat{c}$, it is convenient to perform a standard linearization $\hat{c} = \alpha + \delta\hat{c}$, where the steady-state cavity response $\alpha^2 \sim 1/[\kappa^2 + \Delta^2]$, shown in figure \ref{Fig:LinCoupl}(b), is sensitive to the detuning $\Delta = \omega_{\rm L} - \omega_{\rm c}$ of the laser frequency $\omega_{\rm L}$ from the cavity mode frequency $\omega_{\rm c}$. The response width is given by the cavity decay $\kappa$.
Since the steady-state field  provides the atomic (harmonic) potential, the atomic trap frequency $\omega_{\rm a}$ depends on the field amplitude $\alpha$.
The free Hamiltonian for the setup reads, in a frame rotating with the laser frequency,
\[
H_0 = - \Delta \delta\hat{c}^\dag \delta\hat{c} + \omega_{\rm m} a_{\rm m}^\dag a_{\rm m} +
\omega_{\rm a} (\alpha) a_{\rm a}^\dag a_{\rm a} .
\]
Due to the backaction of the membrane motion on the cavity field, the cavity detuning $\Delta$ is modulated by the membrane motion. The result is a modulation of the field amplitude $\alpha$ (see e.g. the left peak in figure \ref{Fig:LinCoupl}(b)) and consequently the atomic frequency $\omega_{\rm a} (\alpha)$ following the membrane motion.
Thus a single driven cavity mode mediates coupling of the membrane displacement to the atomic motional frequency.
Our goal is a {\it linear} coupling of the atomic {\it displacement} to the membrane motion, via the cavity.

The basic idea behind the linear coupling is to combine two driven fields $\hat{c}_1$ and $\hat{c}_2$, detuned on different sides of their respective resonances, $\Delta \equiv \Delta_1 = - \Delta_2$, as shown in figure \ref{Fig:LinCoupl}(b).
Figure \ref{Fig:LinCoupl}(c) shows the atom in the potential combined of the two cavity fields, when the membrane is in equilibrium position.
When the detunings are shifted due to the membrane motion (dashed line in figure \ref{Fig:LinCoupl}(b)), the
amplitudes of the two fields respond oppositely: when $\alpha_1$ decreases, $\alpha_2$ increases, and vice versa, as
shown in figure \ref{Fig:LinCoupl}(d).
The combined effect is a trap displacement which follows the membrane motion.
At this point it is clear that resonance is obtained when the membrane frequency (and thus the displacement modulation) equals the trap frequency, $\omega_{\rm m} = \omega_{\rm a} $.

Technically, we write the respective couplings of membrane and atom motion to the cavity modes,
\begin{eqnarray}
H_{\rm int} &=& \left( g_{\rm c,m}(a_{\rm m}^\dag + a_{\rm m} ) + g_{\rm c,a} (a_{\rm a}^\dag + a_{\rm a})\right)(\delta\hat{c}_1 + \delta\hat{c}_1^\dag) \nonumber \\
&+& \left( g_{\rm c,m}(a_{\rm m}^\dag + a_{\rm m} ) - g_{\rm c,a} (a_{\rm a}^\dag + a_{\rm a})\right)(\delta\hat{c}_2 + \delta\hat{c}_2^\dag).
\end{eqnarray}
Adiabatic elimination of the cavity fields, under assumption of large detuning $\Delta$ on the scale of the cavity couplings $g_{\rm c,m},g_{\rm c,a}$, yields an effective {\it linear} cavity-mediated membrane-atom interaction. In the crude limit $|\Delta |\gg \omega_{\rm m},\kappa$ the effective coupling term reads,
\[
H_{\rm m,a} = - {4 g_{\rm c,m} g_{\rm c,a} \over \Delta}
 (a_{\rm m}^\dag + a_{\rm m} )(a_{\rm a}^\dag + a_{\rm a}).
\]
Remarkably, for this setup, the strong-coupling regime is reachable with state-of-the-art experimental parameters
even for a single atom \cite{HammererArxiv2009}.

\subsection{Establishing EPR channels between nanomechanics and atomic ensembles}

\begin{figure}[hb]
\begin{centering}
\includegraphics[width=0.95\textwidth]{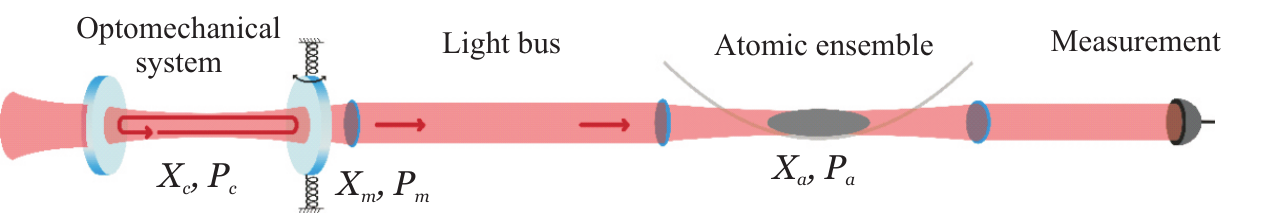}
\caption{Cascaded setup for creating EPR-type entanglement between the mirror motion in an optomechanical system
and the collective spin of an atomic ensemble.}
\label{fig:EPRsetup}
\end{centering}
\end{figure}

After a specific example on a hybrid system with only a {\it single} atom, let us move over to the intriguing world of atomic ensembles and continuous variable theory. Namely, the toolbox for continuous variable quantum protocols has been implemented with atomic ensembles:
EPR-type entanglement generation between ensembles \cite{JulsgaardNature2001},
state swapping from light to ensemble \cite{JulsgaardNature2004}
and teleportation protocols \cite{ShersonNature2006}.
The collective spin of a highly polarized atomic ensemble can be described by continuous variables obeying $[X_{\rm a},P_{\rm a}]= i$. In fact, light and mechanical motion can be similarly described, $[X_{\rm m,c},P_{\rm m,c}]= i$, thus it is possible to adapt the continuous variable toolbox to a setup including an atomic ensemble and an optomechanical system, where the
mechanical resonator is either a movable Fabry-Perot cavity mirror \cite{GiganNature2006}, or
a dispersive membrane in a rigid cavity \cite{ThompsonNature2008}.
An optical quantum bus provides an interface for the nanomechanical system to the quantum toolbox already available for control of atomic ensembles, including high fidelity preparation and read-out via light, and robust high fidelity storage of quantum states.

The proposal of \cite{HammererPRL2009} is to create robust entanglement between collective atomic spin variables
and the motion of a nanomechanical resonator, exhibiting reduced EPR variance of correlations,
\[
\Delta_{\rm EPR} = \Delta (X_{\rm m} + X_{\rm a})^2  + \Delta (P_{\rm m} - P_{\rm a})^2 < 2 .
\]
The generated entanglement can serve as basis for teleporting quantum states of collective spin onto a
nanomechanical system, and further implies the intriguing possibility to cool a mechanical resonator by teleporting the ground state onto it.

In this setup, atoms and nanomechanical mode couple to light via radiation pressure and Faraday coupling, respectively,
\[
H = {\omega_{\rm m} \over 2} (X_{\rm m}^2 + P_{\rm m}^2) + {\Omega \over 2} (X_{\rm a}^2 + P_{\rm a}^2)
+ g (X_{\rm m} + X_{\rm a}) X_{\rm c}
\]
assuming couplings tuned to equal strength $g$.
The proposed method for creating entanglement, is to perform a QND measurement of the commuting observables $X_{\rm m} + X_{\rm a}$ and $P_{\rm m} - P_{\rm a}$ which {\it projects} the atomic-nanomechanical system into a state with reduced variances.
Crucial is that the atomic ensemble is polarized in the energetically higher lying state of the two ground states, such that the collective atomic variables describe a harmonic oscillator with negative mass,
$\Omega = - \omega_{\rm m}$. In that case, both $X_{\rm m} + X_{\rm a}$ and $P_{\rm m} - P_{\rm a}$ are conserved quantities, thus allowing QND measurement of both observables. Moreover, the interaction Hamiltonian $H_{\rm I}$,
\[
H_{\rm I} = g [\cos(\omega_{\rm m}t)X_{\rm c}(X_{\rm m} + X_{\rm a}) + \sin(\omega_{\rm m}t)X_{\rm c}(P_{\rm m} - P_{\rm a})]
\]
shows that for a fast decaying cavity mode, the relevant observables $X_{\rm m} + X_{\rm a}$ and $P_{\rm m} - P_{\rm a}$ are linked to the cosine and sine components of the output light, respectively, and thus are detectable by homodyne measurement. A cascaded setup, as shown in figure \ref{fig:EPRsetup}, where the output light of a laser driven optomechanical system is fed into the
atomic ensemble setup, followed by homodyne measurement, allows for distant EPR correlations between systems acting in different environments. A nice feature of this proposal is that it does not require ground-state cooling of the nanomechanical resonator. The reason is that in the limit of strong coupling, a QND measurement realizes a projective von Neumann measurement which collapses the system into a pure, entangled state irrespective of initial conditions.

\section{Cavity QED and quantum networks with microtoroidal cavities}
\label{Sec:network}

Finally, we comment briefly on promising recent developments in combining atomic cavity QED involving solid state high-Q cavities. This is of interest in the context of building quantum networks. A quantum network is defined as a collection of nodes coupled by channels. The nodes store and process quantum information locally, while channels provide quantum interconnects between the various nodes, e.g. in the sense of long distance quantum communication. In an AMO context the realization of such a network is provided by optical cavity QED, where atoms representing the quantum memory are stored in cavities. These cavities provide a local quantum data bus via exchange of photons, but also an interface with flying photonic qubits.

An example of a hybrid AMO  solid-state system of such a network has been discussed by Kimble and collaborators \cite{KimbleNature2008}. In this  microtoroidal cavity-QED setup a whispering-gallery cavity mode couples strongly to individual atoms \cite{AokiNature2006} and can connect with high efficiency to connecting optical fibre quantum channels.
Furthermore, a single atom within the resonator can dynamically control the cavity output, thus making possible
a photon turnstile mechanism for regulated transport of photons one by one \cite{DayanScience2008}. Strong coupling was also demonstrated for a cavity-QED setup with an artificial atom quantum dot embedded in a microdisk \cite{SrinivasanNature2007}. A similar setup using atom-chip technology was demonstrated in \cite{ColombeNature2007}, with the field of an onchip fiber-based Fabry-Perot cavity showing strong coupling on the single-atom level to an ensemble quantum memory in form of a BEC.


\section{Conclusions}

During the last fifteen years we have seen proposals and successful small scale realizations for implementing quantum information processing, in the context of atomic, molecular and optical physics, but also in solid state physics.  With the maturing of the field of experimental quantum information, it seems timely to consider  hybrid quantum systems involving atomic and solid state elements with the goal of combining the advantages of the various systems in compatible experimental setups. Here we have reviewed several promising new directions of building such hybrid quantum systems involving atomic and solid state elements. In particular, we have focused on quantum interfaces and interconnects between AMO and solid-state quantum circuits mediated by photons in the optical and microwave domain in CQED and free space setups, as well as a phonon based quantum data bus with nanomechanical elements. We believe that these ideas are a promising new route in the next generation development of experimental quantum information processing.

\ack

We acknowledge support by the  Austrian Science Fund through SFB FOQUS, by the Institute for Quantum Optics and Quantum Information, and by the European Union through project EuroSQIP. P.R. and M.L. acknowledge support by ITAMP, NSF and the Packard Foundation.


\end{document}